\documentclass[preprint,aps]{revtex4}

\usepackage{graphicx}
\usepackage{dcolumn}
\usepackage{bm}

\begin{document}


\title{Recursive approach to the calculation of
a many-body basis in discrete electronic nanosystems}

\author{K. M. Indlekofer}
\email{m.indlekofer@fz-juelich.de} \affiliation{Center of Nanoelectronic Systems
for Information Technology (CNI), IBN-1, Research Centre J\"ulich GmbH,
D-52425 J\"ulich, Germany}

\author{R. N\'{e}meth}
\affiliation{Institute of Electrical Engineering,
Slovak Academy of Sciences, D\'{u}bravsk\'{a} 9,
SK-84104 Bratislava, Slovak Republic}

\date{\today}

\begin{abstract}
In this article, we describe a recursive method to construct a
subset of relevant Slater-determinants for the use in many-body diagonalization
schemes that will be employed in our forthcoming papers on the
simulation of excited many-body states in discrete electronic nanosystems.
The algorithm is intended for the realistic simulation of nanodevices
which typically requires the consideration of a large number
of single-particle basis states (typ. 256) and the calculation of a
sufficient number (typ. a few 1000) of relevant excited many-body states.
\end{abstract}

\maketitle

\section{Introduction}

Realistic microscopic modeling of application-relevant electronic
nanostructures typically requires the consideration of 100-1000
single-particle basis states (orbitals).
Normally, due to the lack symmetries, such
systems are inhomogeneous and anisotropic and furthermore may exhibit a
long-range interaction.
With a suitable unitary transformation
and a subsequent division of the single-particle Hilbert space,
the number of relevant single-particle states
that require a many-body treatment of the Coulomb interaction
typically can be reduced to $\leq$256.
Even with this reduction, the resulting
Fock space dimension obviously grows beyond practical numerical
limits, at least for a full exact diagonalization (full-CI). In
order to address this problem one can take the path to consider a
highly idealized model
(such as Anderson- or Hubbard-like models \cite{anderson,hubbard}),
mapping the actual nanosystem to a reduced (albeit non-trivial)
model system with a small number of effective parameters. Such an
approach can be justified in the sense that only a small number of
degrees of freedom typically are relevant for the investigated fundamental
effects. Nevertheless, these relevant degrees of freedom and their
associated effective parameters in general are non-linear functions
of the actual experimentally accessible parameters (such as gate
voltages and general external electromagnetic fields) and not known
a-priori. As an advantage however, due to the simplified structure
of the model Hamiltonians, one is able to understand the physics of
such idealized models in detail (in particular, the possibility to
correctly include continuous semi-infinite contacts). However, in
order to determine effective parameters (as a function of external
quantities and material properties) and to obtain a more
realistic description of the nanosystem (in particular for the
simulation of application-relevant structures), methods which
incorporate more microscopic details are mandatory. While ab-initio
methods in principle provide this kind of information, they often
involve approximation schemes (such as mean-field approximations)
that are not suitable for the description of many-body
correlations which certainly cannot be neglected in nanosystems with
quantum effects. Nevertheless, ab-initio basis functions and
matrix elements might serve as input parameters for subsequent
many-body approaches.
In this context, very sophisticated methods have
been developed in the field of many-body physics
(see Refs.~\cite{reimann,wilson_nrg,hofstetter,wetterich,morris,white_dmrg,schollwoeck,schoeller,white_cd,wegner,glazek,raedt,foulkes,ceperley,filinov})
and quantum chemistry (see Refs.~\cite{szabo,mueller,dronskowski,sherrill,szalay,schautz}).
Typically, these approaches are competitive only for specialized systems or
conditions, such as: groundstate,
equilibrium or close to equilibrium,
idealized model interaction or periodic systems,
limited-particle excitations,
small single-particle basis, restriction to systems
with special symmetries (such as spin and total angular momentum, leading to
block-diagonal Hamiltonians),
restriction to Fock-Darwin single-particle basis
(or even 2D-harmonic potential only).

In the following, we discuss a recursive scheme which is able
to describe a finite (or quasi-isolated) nanosystem with a large number
of single-particle basis states (typ. 256) and to calculate a
sufficient number (typ. a few 1000) of relevant excited many-body states.
Note that the states we are interested in
are not necessarily close to the groundstate.
This approach consists of two steps: (i) Recursive construction of a
large set of relevant Slater-determinants ("bucket brigade
algorithm") as discussed in this paper.
(ii) Many-body calculations within the resulting Fock-subspace.

\section{Recursive construction of relevant Slater-determinants}

For an arbitrarily given single-particle orthonormal (ON) basis $B=\{\phi_j\}$ the
exact many-body Hamiltonian of the finite (i.e., discrete) system
in second quantization reads as
\begin{equation}
H=
\sum_{j=0}^{\infty} \sum_{k=0}^{\infty} \epsilon_{jk} c^{\dagger}_{j} c_{k}
+
\sum_{k=1}^{\infty} \sum_{j=0}^{k-1} \sum_{l=1}^{\infty} \sum_{m=0}^{l-1}
V_{jklm} c^{\dagger}_{j} c^{\dagger}_{k} c_{l} c_{m},
\end{equation}
where $\epsilon_{jk}$ and $V_{jklm}$ denote single-particle and
Coulomb two-particle matrix elements of the total system
Hamiltonian, and $c_j$ is the annihilation operator for single-particle
state $j$.

\begin{figure}
\centering
\includegraphics[width=5in]{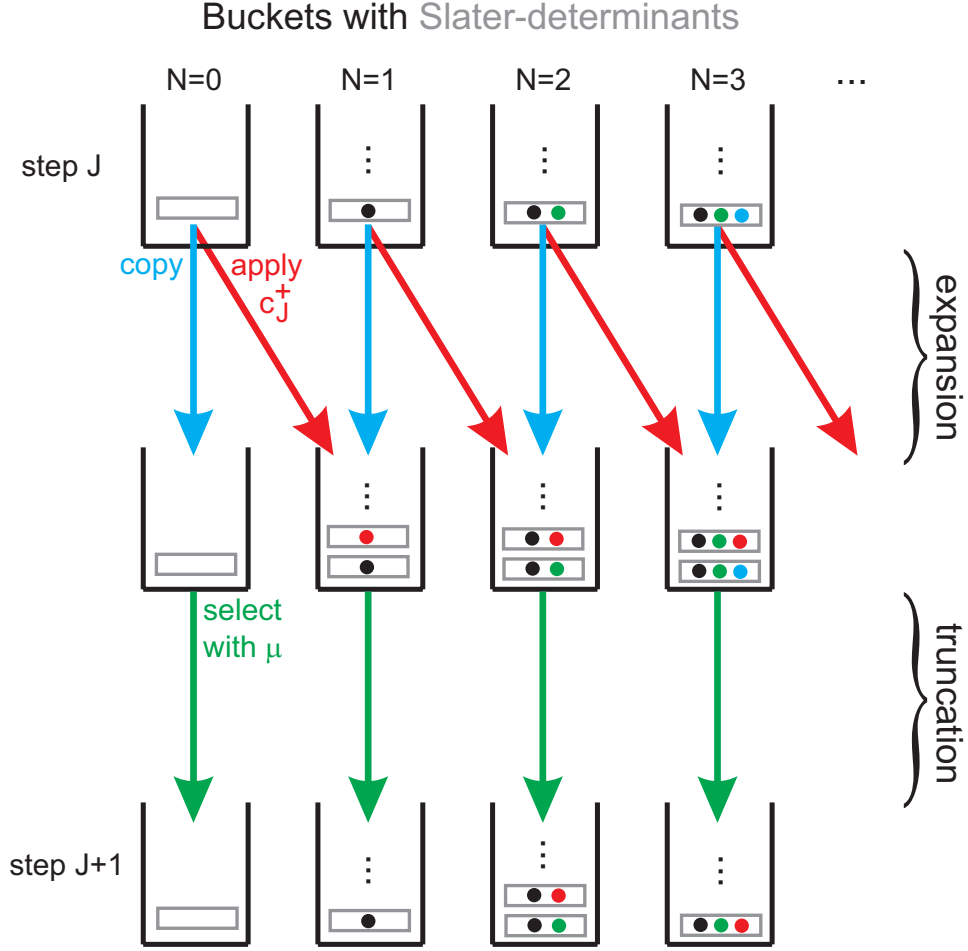}
\caption{Visualization of the recursion step $J\to J+1$:
for a considered bucket $N$ copy all existing states,
add new states by applying $c^{\dagger}_J$ to all existing states from previous bucket ($N-1$)
and truncate with the help of the measure $\mu$.
Filled circles represent single-particle states and $N$ denotes the particle number in the bucket.
(The shown truncation is just an example.)}
\label{fig1}
\end{figure}

Let $N_{max}$ be the maximum number of single-particle states that
shall be considered. Hence, we consider a subspace spanned by
$B_{N_{max}}=\{\phi_j\mid j=0,...,N_{max}-1\}$. The recursion
procedure can be formulated in terms of a sequence of sets $S_{J,N}$
of Slater-determinants. Here, $J=0,...,N_{max}$ denotes the
recursion step number and $N=0,...,N_{max}$ is the particle number.
$S_{J,N}$ is defined via the following ``bucket brigade'' recursion
as visualized in Fig.~\ref{fig1}:
\begin{itemize}
\item
{\it Start (initial conditions)}\\
For $J=0,...,N_{max}: S_{J,0}=\{|0\rangle\}$,\\
for $N=1,...,N_{max}: S_{0,N}=\emptyset$.
\item
{\it Recursion step} $J\to J+1$\\
For $J=0,...,N_{max}-1$ and $N=1,...,N_{max}:$
\begin{itemize}
\item
``Expansion''\\
Let $\hat{S}_N=S_{J,N}\cup\{c^{\dagger}_J v \mid v \in S_{J,N-1}\}$.
\item
``Truncation''\\
If $\mbox{card}(\hat{S}_N)\leq D_{N}$: $S_{J+1,N}=\hat{S}_N$.\\
Else: choose $S_{J+1,N}\subset \hat{S}_N$ such that\\
\phantom{........} $\mbox{card}(S_{J+1,N})=D_N$\\
\phantom{........} and $\forall v \in S_{J+1,N}: \forall w \in \hat{S}_N\backslash S_{J+1,N}:
\mu(v)\leq \mu(w)$.
\end{itemize}
\end{itemize}
Here, $|0\rangle$ denotes the Fock space vacuum state,
$\mbox{card}(S)$ is the number of elements in set $S$, and
$D_N$ (with $N=0,...,N_{max}$) are given integers which determine the maximum
number of states in $S_{J,N}$ (for all $J$). Note that the maximum
number of Slater-determinants for $N$ particles reads as
\begin{equation}
D_{max}(N)=\frac{N_{max}!}{(N_{max}-N)!~N!}.
\end{equation}
Hence, we must have $D_N\leq D_{max}(N)$.
Furthermore, $\mu$ is a suitable real ``measure'' for the selection of
Slater-determinants.
For a normalized Slater-determinant $v$, we assume a measure
of the form $\mu(v)=\langle v|M|v\rangle$ with a (w.o.l.g. diagonal)
many-body operator $M$. The latter can be expanded in a series of
$n$-particle product terms
\begin{equation}
M=\sum_{j_1} M^{(1)}_{j_1} N_{j_1}
+\sum_{j_1<j_2} M^{(2)}_{j_1 j_2} N_{j_1}N_{j_2}
+\ldots,
\end{equation}
with $N_j\equiv c^{\dagger}_j c_j$ for single-particle state $j$ an
real coefficients $M^{(n)}$, which represent conditional
$n$-particle selection weights. In the simplest case, one can choose
the diagonal elements of the Hamiltonian, that is
\begin{eqnarray}
M^{(1)}_{j}&=&\epsilon_{jj},
\\\nonumber
M^{(2)}_{jk}&=&V_{jkkj}-V_{jkjk},
\\\nonumber
M^{(n>2)}&=&0.
\end{eqnarray}
(Note that this choice might not provide a unique set if some Slater-determinants
are degenerate.) Optionally, the final truncation
condition can be made even more selective, choosing for example only
such Slater-determinants which have energies that are within certain
energy intervals (such as ``bands''). From this construction it is
obvious that the set $S_{J,N}$ contains only Slater-determinants
with a total number of $N$ particles. As a physical interpretation,
with each recursion step $J\to J+1$ the set $S_{J+1,N}$
consists of the lowest $D_N$ states from the combination of
$S_{J,N}$ and all states of $S_{J,N-1}$ with an extra particle
created in the single-particle state $j=J$.
To some extent, the
described procedure resembles the NRG recursion scheme, however,
lacking the diagonalization steps at this stage (see next
paragraph).

If we are solely interested in calculating expectation values of
observables for a fixed particle number $N_0$, it is sufficient to
calculate $S_{J,N}$ for $N=0,...,N_0$ only and to consider the last
set $S_{N_{max},N_0}$ in the following . However, if $n$-point
Green's functions (i.e., contour-ordered correlation functions of
$n/2$ creation and annihilation operators at $n$ points in time)
have to be calculated as well, we typically need information about
all Fock subspaces of particle numbers
$\tilde{N}_0=N_0-n/2,...,N_0+n/2$. Obviously, $n$ is restricted to
$0\leq n/2\leq \min(N_0,N_{max}-N_0)$. Since the single-particle
subspace basis $B_{N_{max}}$ might be optimized only for states with
$N_0$ particles, it could become sub-optimal for $\tilde{N}_0\neq
N_0$. In general, we thus could consider an individual subspace
basis $B^{\tilde{N}_0}_{N_{max}}$ for each $\tilde{N}_0$ and an
individual $S_{J,N}[B^{\tilde{N}_0}_{N_{max}}]$.

From the obtained $S_{N_{max},N}$, we can now define a finite Fock
subspace
\begin{equation}
F_0=\mbox{span}(\tilde{S})
\end{equation}
spanned by the many-body basis $\tilde{S}$ of ``relevant'' Slater-determinants
\begin{equation}
\tilde{S}=\bigcup_{N=N_0-\frac{n}{2}}^{N_0+\frac{n}{2}}
S_{N_{max},N}
\end{equation}
and with dimension
\begin{equation}
\dim(F_0)=\sum_{N=N_0-\frac{n}{2}}^{N_0+\frac{n}{2}}
\mbox{card}(S_{N_{max},N})
\leq\sum_{N=N_0-\frac{n}{2}}^{N_0+\frac{n}{2}} D_{N}.
\end{equation}
Note that the states of $S_{N_{max},N}$ for different $N$ are
orthogonal to each other since they obviously have different
particle numbers. By construction, the states within $S_{N_{max},N}$
for a fixed $N$ are orthonormalized. Hence, $\tilde{S}$ is a
many-body ON-basis of $F_0$. In turn, a corresponding finite
restricted Hamiltonian can be defined as
\begin{equation}
H_0=H\mid_{F_0},
\end{equation}
which acts within $F_0$. As the final step, $H_0$ is diagonalized
numerically in a matrix representation with respect to the basis
$\tilde{S}$. Since $H$ conserves the particle number, $H$ is
block-diagonal within individual $\mbox{span}(S_{N_{max},N})$. With
the help of the eigenvectors and eigenvalues of $H_0$ one can
calculate the time evolution of the isolated system and can
construct stationary statistical operators for the calculation of
expectation values.

By considering only those Slater-determinants that are chosen as "relevant"
by the described algorithm, two controlled approximations are made:
First, the number of considered single-particle states $N_{max}$ is
finite. Equivalently, the recursion step is performed only a finite
number of times $N_{max}$. Second, the maximum number $D_N$ of
Slater-determinants in each set $S_{J,N}$ can be smaller than
$D_{max}(N)$. In the limit $N_{max}\to\infty$ and $D_N\to\infty$
($N=0,...,N_0$), the algorithm becomes exact.

The described algorithm typically becomes advantageous if the
required maximum number of single-particle states $N_{max}$ and the
number of particles $N_0$ provides a number $D_{max}(N_0)$ of
Slater-determinants that lies beyond the limit for which the resulting
problem can be fully diagonalized, in particular for strongly
inhomogeneous systems with minimal symmetries.
Typical application-relevant values are
$N_{max}=32-256$, $N_0=0-32$, and $D_N=1024-16384$. For such a
regime, NRG-based approaches might become impractical since the
storage requirements of matrices for the given $N_{max}$ and $D_N$
grow beyond realistic limits. In principle, the described algorithm
is suitable to provide a basis for many-body states up to any energy scale
since all single-particle states are scanned systematically.
It can be used in particular for many-body problems where a single set of
Slater-determinants has to be determined
which is relevant for a whole energy interval of many-body eigenstates
(not only the groundstate or a particular excited state).
An implementation on parallel computer architectures is possible.

\section{Optimization}

Within this algorithm, two means of (a-posteriori)
optimization exist: (i) Choice of the single-particle basis and in
particular its order. (ii) Choice of the selection criterion $\mu$
of Slater-determinants.

As for (i), one could solve an effective single-particle mean-field
problem derived from $H$ for a given particle number $N_0$ and the
given occupation conditions (e.g., Hartree-Fock, LDA-DFT \cite{kohn_dft,perdew_dft}).
The resulting single-particle ON eigenbasis of this problem could be
chosen for $B$. Alternatively, one can define an outer
self-consistency loop for the described algorithm. Here, the
many-body state (preparation) of the system is decribed by a
suitable many-body statistical operator $\rho[H]$ which is expressed
in terms of the calculated basis $\tilde{S}$ and the associated
eigenvalues for a given boundary condition (such as maximum entropy
for thermodynamical equilibrium, non-equilibrium injection
conditions, or the selection of a certain excited many-body state).
In turn, the single-particle eigenstates of the resulting
(transposed) single-particle density-matrix $\hat{\rho}$, which can
be calculated via
\begin{eqnarray}
\hat{\rho}_{kj}&=& \frac{1}{i}~ G^<_{jk}(t=0)
\\\nonumber
\mbox{with}\quad
G^<_{jk}(t)&=&
i~ \mbox{Tr}\left( \rho~ c^{\dagger}_{k}(0) ~c_{j}(t) \right),
\end{eqnarray}
are used as a new single-particle basis $B_{new}$ for a repeated
many-body diagonalization, employing $\tilde{S}[B_{new}]$. Such an
approach resembles the MCSCF formalism \cite{szabo} (where these single-particle
states are referred to as ``natural orbitals''), which is used
typically for the calculation of the many-body groundstate.
The correct choice of the single-particle basis is essential for
any finite expansion of many-body states in Slater-determinants to be efficient.

As for (ii), the measure $\mu$ can also be chosen adaptively from
the many-body diagonalization. Here, the coefficients $M^{(n)}$ can
be constructed from the probability weights of conditional
$n$-particle projection amplitudes of the calculated many-body
statistical operator $\rho$. Implemented as a repeated
selection+diagonalization, this defines a self-consistent
optimization scheme for $\mu$.

In a numerical implementation, the calculation (or transformation)
and storage of the Coulomb matrix elements $V_{jklm}$ might become a
challenge ($[(N_{max}-1)N_{max}/2]^2$ independent index
combinations). Here, symmetries of $V$ have to be employed in
combination with possible reduction and approximation schemes.

\section{Summary}

We have discussed a recursive method to construct a
subset of relevant Slater-determinants for the use in many-body diagonalization
schemes in order to simulate many-body effects of realistic nanodevices
with Coulomb interaction. The system is assumed to be finite.
The described method becomes advantageous if a large number
of single-particle basis states (typ. 256) is required where full diagonalization
schemes become impractical.
By use of a recursive method, a reduced number (typ. a few 1000) of relevant
many-body basis states (Slater-determinants) is generated, systematically
scanning all given single-particle basis states.

\end{document}